\documentclass[bst/sn-mathphys,Numbered]{sn-jnl}

\usepackage{graphicx}%
\usepackage{multirow}%
\usepackage{amsmath,amssymb,amsfonts}%
\usepackage{amsthm}%
\usepackage{mathrsfs}%
\usepackage[title]{appendix}%
\usepackage{xcolor}%
\usepackage{textcomp}%
\usepackage{manyfoot}%
\usepackage{booktabs}%
\usepackage{algorithm}%
\usepackage{algorithmicx}%
\usepackage{algpseudocode}%
\usepackage{listings}%
\usepackage[square,numbers]{natbib}
\usepackage{ulem}
\usepackage{mathabx}

\usepackage{tabularx} 
\usepackage{makecell} 
\usepackage{booktabs} 
\newcolumntype{Y}{>{\centering\arraybackslash}X} 
\newcolumntype{R}[1]{>{\raggedleft\arraybackslash}p{#1}} 

\def\mb#1{\textcolor{blue}{#1}}

\begin{document}


\title{Different arrival times of CM and CI-like bodies from the outer Solar System to the asteroid belt}







\author*[1]{\fnm{Sarah E.} \sur{Anderson}}\email{sarah.anderson@lam.fr}

\author[1]{\fnm{Pierre} \sur{Vernazza}}\email{pierre.vernazza@lam.fr}

\author[2]{\fnm{Miroslav} \sur{Bro\v z}}\email{mira@sirrah.troja.mff.cuni.cz}

\equalcont{These authors contributed equally to this work.}

\affil*[1]{\orgdiv{LAM, Laboratoire d’Astrophysique de Marseille}, \orgname{Aix Marseille University}, \orgaddress{\street{38 rue Frédéric Joliot-Curie}, \city{Marseille}, \postcode{13013}, \country{France}}}

\affil[2]{\orgname{Charles University, Faculty of Mathematics and Physics}, \orgdiv{Institute of Astronomy}, \orgaddress{\street{V Holešovičkách 2}, \city{Praha}, \postcode{CZ-18200}, \country{Czech Republic}}}


\abstract{

Understanding the provenance of  CI and CM chondrites, the most primitive materials in our meteorite collections, is critical for shedding light on the Solar System's early evolution \citep{Morbidelli2005, Bottke2006, Levison2009, Walsh2011, Raymond2017a} and contextualizing findings from recent sample return missions \citep{Marsset_2024arXiv240308548M,Broz_2024arXiv240308552B}.
Here we show that the parent bodies of CM chondrites originate from the Saturn formation region, whereas those of CI chondrites originate essentially from the primordial trans-Uranian disk. Using N-body simulations to investigate the effect of giant planet growth and inward Type-I migration \citep{Nesvorn2015, Deienno2017}, along with the current observed distribution of CM, CI, and comet-like (P types) bodies in the asteroid belt \citep{DeMeo2013}, we demonstrate that CI and CM-like bodies must have been implanted at different times in the belt, while CI and comet-like bodies were implanted at the same time. These different implantation periods are imposed by the fact that the gas disk profile entirely governs the radial distribution of bodies implanted by aerodynamic drag in the asteroid belt.
A preferred location coincides with the inner edge of a `gap' opened by Jupiter. Saturn's growth likely drove the migration of CM-like bodies, whereas CI and comet-like bodies were transported at a later stage, during the outward migration of Uranus and Neptune driven by remaining planetesimals. Since CM chondrites are chondrule-rich,
it follows that chondrule formation occurred mostly inward of the ice giant's formation zone ($\leq$10\,au).
A by-product of our simulations is that only CM-like (not CI-like) bodies contributed
to the water budget of the telluric planets.
}




\maketitle





To date, the vast majority of meteorites ($\geq$99\%) originate from asteroids in orbits between Mars and Jupiter \citep{Marsset_2024arXiv240308548M,Broz_2024arXiv240308552B}. The challenge of aligning the large chemical, compositional, and isotopic diversity of meteorites with a model of in-situ formation in the asteroid belt has long eluded consensus. Dynamical models have significantly shifted this paradigm, suggesting that main-belt asteroids --- hence meteorites --- formed over a large range of heliocentric distances \citep{Morbidelli2005, Bottke2006, Levison2009, Walsh2011, Raymond2017a}, from the terrestrial planet formation region to the Kuiper Belt. This view has been progressively supported by meteorite measurements revealing first a dichotomy between carbonaceous and non-carbonaceous chondrites \citep{Warren2011, Budde2016, Kruijer2017} and more recently a further dichotomy among carbonaceous chondrites (CCs), between CIs and the remaining CCs \citep{Hopp2022}. Whereas specific formation locations have been proposed for the different CC groups (most CCs: Jupiter-Saturn formation region, CI chondrites: Neptune formation region; \citep{Nesvorn2024}) no single dynamical model has been able to correlate these putative formation locations with the current observed compositional distribution of the asteroid belt. Apart from a general consensus that most CCs formed beyond Jupiter \textemdash and even this remains to be unambiguously demonstrated \textemdash the origin of the diversity among CC groups, including their various formation locations, remains a mystery. \\

\begin{figure}
    \centering
    \includegraphics[width=\textwidth]{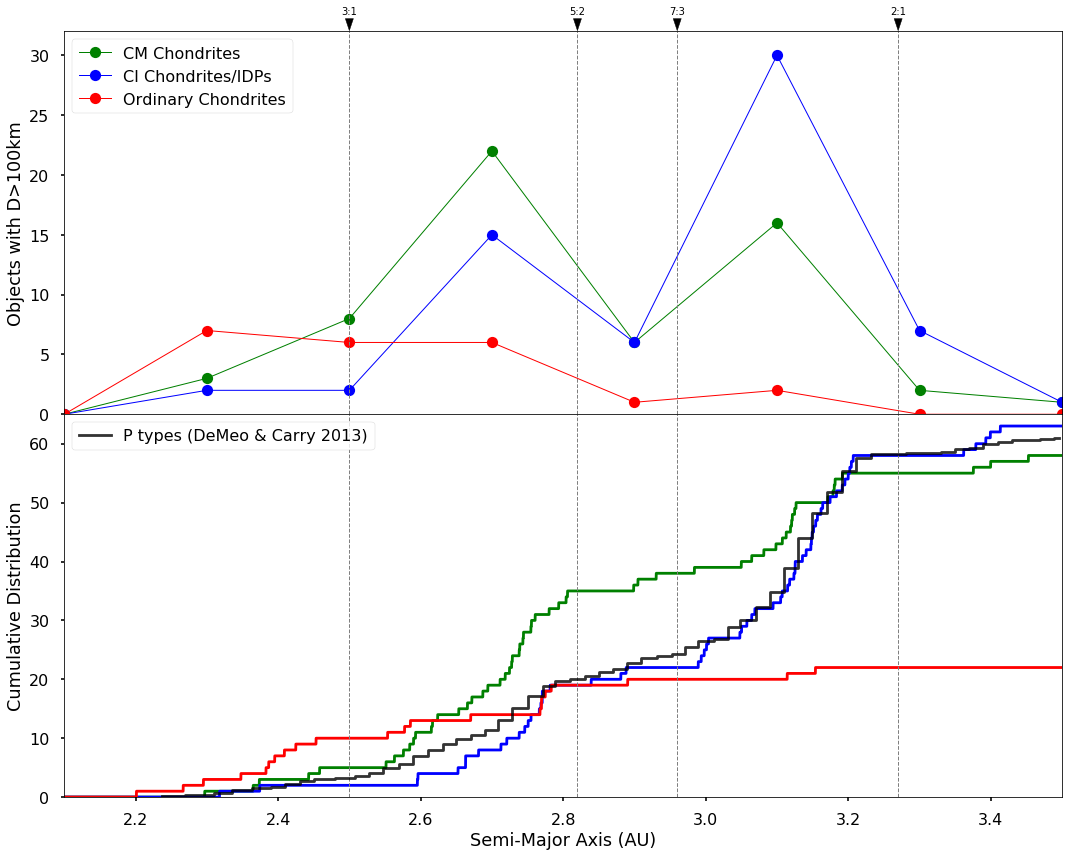}
    \caption{
    {\bf Observed distributions of ${>}100$\,km main-belt asteroids with distinct spectral classifications (CM, CI/IDP, B, S) are systematically different.}
    Top: The number of asteroids classified as chondrule-rich CM (green), chondrule-poor CI/IDP and B-type possessing an albedo below 0.1 (blue), and S-type ordinary-chondrite asteroids (red). We prioritize numerical frequency over mass distribution, a decision driven by the fact that Ceres alone constitutes approximately 90\% of the asteroid belt's total mass. To ensure statistical integrity in our evaluations, asteroid families are considered as singular entities, thereby preventing potential overcounts. Bottom: Cumulative distributions of the semimajor axis $N({\leq}a)$ for the same classes of asteroids. For P-types, all ${>}30$\,km asteroids were included \citep{DeMeo2013}, as there are only a dozen P-types ${>}100$\,km. These are normalized to the number of CI/IDPs at 3.2\,au. The positions of major mean-motion resonances with Jupiter (3:1, 5:2, 7:3, 2:1) are indicated by dotted lines. Whereas the distributions of CMs and CI/IDPs are different, CI/IDPs and P-types seem to be compatible according to the Kolmogorov--Smirnov test.
    }
    \label{fig:CMCI}
\end{figure}

Today, two meteorite classes are over-represented among CC-like asteroids namely CM chondrites (Ch, Cgh-types \citep{Vernazza2016}) and CI chondrites (B, C, Cb, and Cg-types \citep{DeMeo2022, Ito2022, Hopp2022, Yada2022, Yokoyama2023}). The parent bodies of these two meteorite classes represent more than 50\% of the mass of the asteroid belt (not counting Ceres) \citep{DeMeo2013}. In contrast, the parent bodies of the remaining CC classes (CO, CV, CK) represent a minority ($\leq$1\%) of all main belt asteroids \citep{DeMeo2013}. The radial distribution of CM- and CI-like bodies in the asteroid belt appear very different (Fig.~\ref{fig:CMCI}), with CM-like bodies displaying a rather Gaussian, symmetric profile, whereas CI-like bodies display a markedly asymmetric profile. Of great interest, the distribution of CI-like bodies is remarkably similar to that of comet-like P-type asteroids (Fig.~\ref{fig:CMCI}), suggesting a common origin. A common link between these two asteroid compositions is not new as they were already linked based on spectroscopic and density measurements \citep{Vernazza2015, Vernazza2021}, with a proposed link to interplanetary dust particles (IDPs). \\

To constrain the formation location of CM- and CI-like bodies, and shed light on the origin of the isotopic dichotomy among CCs \citep{Hopp2022}, we used an orbital model to investigate the injection of planetesimals into the asteroid belt following giant planet growth. We conducted simulations using \texttt{REBOUND} (ref.~\citep{Rein2012}), consisting of 20,000 test particles each, representing 100-km-sized planetesimals. Our five-planet model was motivated by the works of Nesvorný \citep{Nesvorn2015} and Dienno \citep{Deienno2017} and featured Jupiter, Saturn, an additional ice giant (which is commonly ejected from the solar system), Uranus, and Neptune. Initially, planets were located on low-eccentricity orbits, with Jupiter at 5.4\,au and Saturn at the edge of its gap, at $7.3$\,au. We chose to neglect the influence of telluric planets, which, having relatively small orbits, require more computation time. We placed Neptune at two different locations, namely in a `tight' configuration at 16.2\,au, in a 3:2, 3:2, 3:2, 3:2 resonance chain, or in a `wide' configuration at 20.3\,au, in a 3:2, 3:2, 2:1, 3:2 resonance chain \citep{Deienno2017}. The test particles' semi-major axes were initialized uniformly between 7\,au and 1\,au beyond the orbit of Neptune. We then examined the effect of gas profile ($\Sigma_{\rm g}$), planetary growth timescale ($\tau_\text{growth}$), or the viscosity parameter ($\alpha$) on the distribution of planetesimals implanted in the asteroid belt (see SI for more details). Notably, we investigated three different gas profiles: the traditional canonical model with a radial dependency of $\Sigma_{\rm g} \propto r^{-0.5}$ \citep{Cresswell2008}, the Desch \citep{Desch2018} model, and the Raymond \& Izidoro \citep{Raymond2017} model.\\



\begin{figure}
    \centering
    \includegraphics[width=\textwidth]{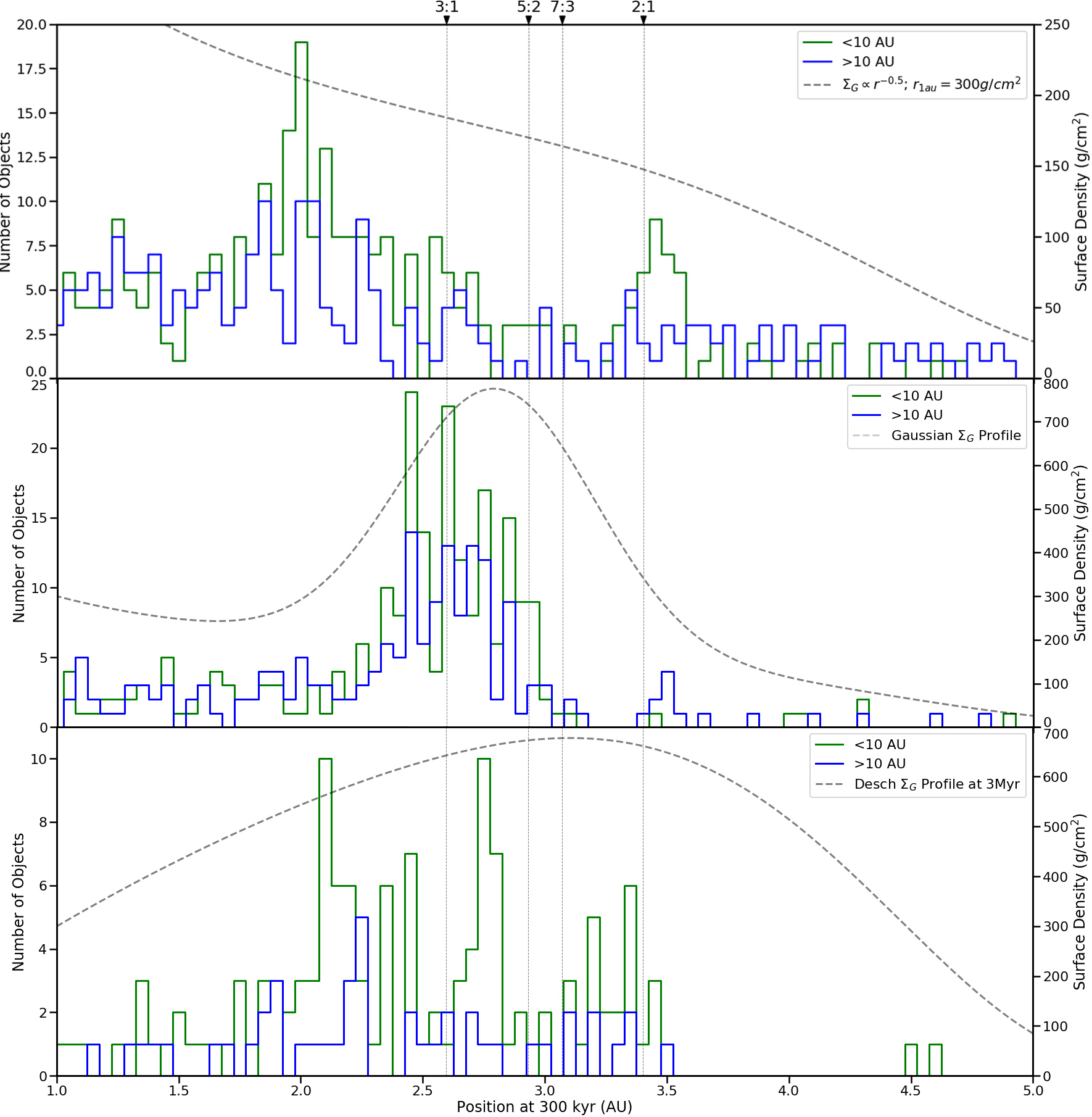}
    \caption{
    {\bf Simulated planetesimals implanted by aerodynamic drag closely follow a gas profile of the protoplanetary disk.}
    The semimajor axis distributions of planetesimals implanted interior to Jupiter (solid lines), along with corresponding gas surface density $\Sigma_{\rm g}(r)$ (dashed lines). The planetesimals are split into those having formed or originated $<10$\,au (green) and $>10$ (blue). Three different profiles are shown:
    the canonical model (top), the model with a pressure bump represented by a Gaussian peak (middle), and the gas profile of Desch at 3 Myr \citep{Desch2018} (bottom). The simulated time span was 300\,kyr. Pressure bumps act as efficient planetesimal traps. As their position subsequently changes, planetesimals are implanted at different locations.}
    \label{fig:gas}
\end{figure}

We find that the most influential factor for the distribution of objects within the asteroid belt proves to be the gas profile of the protoplanetary disk rather than the formation location of small bodies, the giant planet growth timescales, the viscosity parameter of the disk, or the configuration of the planets. The final distribution of objects successfully implanted in the asteroid belt strongly reflects the shape of the gas profile (Fig.~\ref{fig:gas}). We tested our hypothesis further by simulating pressure bumps within the protoplanetary disk, by introducing gas profiles characterized by Gaussian peaks. These artificial constructs highlight the paramount importance of the gas profile in the final distribution of planetesimals interior to Jupiter, regardless of their origin: bodies formed around Saturn will end up with the same radial distribution as those that originate around Neptune. Whatever the gas profile, it is interesting to note a small inward radial shift of implanted bodies relative to the gas distribution (Fig. \ref{fig:gas}).
\\

\begin{figure}
\centering
\includegraphics[width=\textwidth]{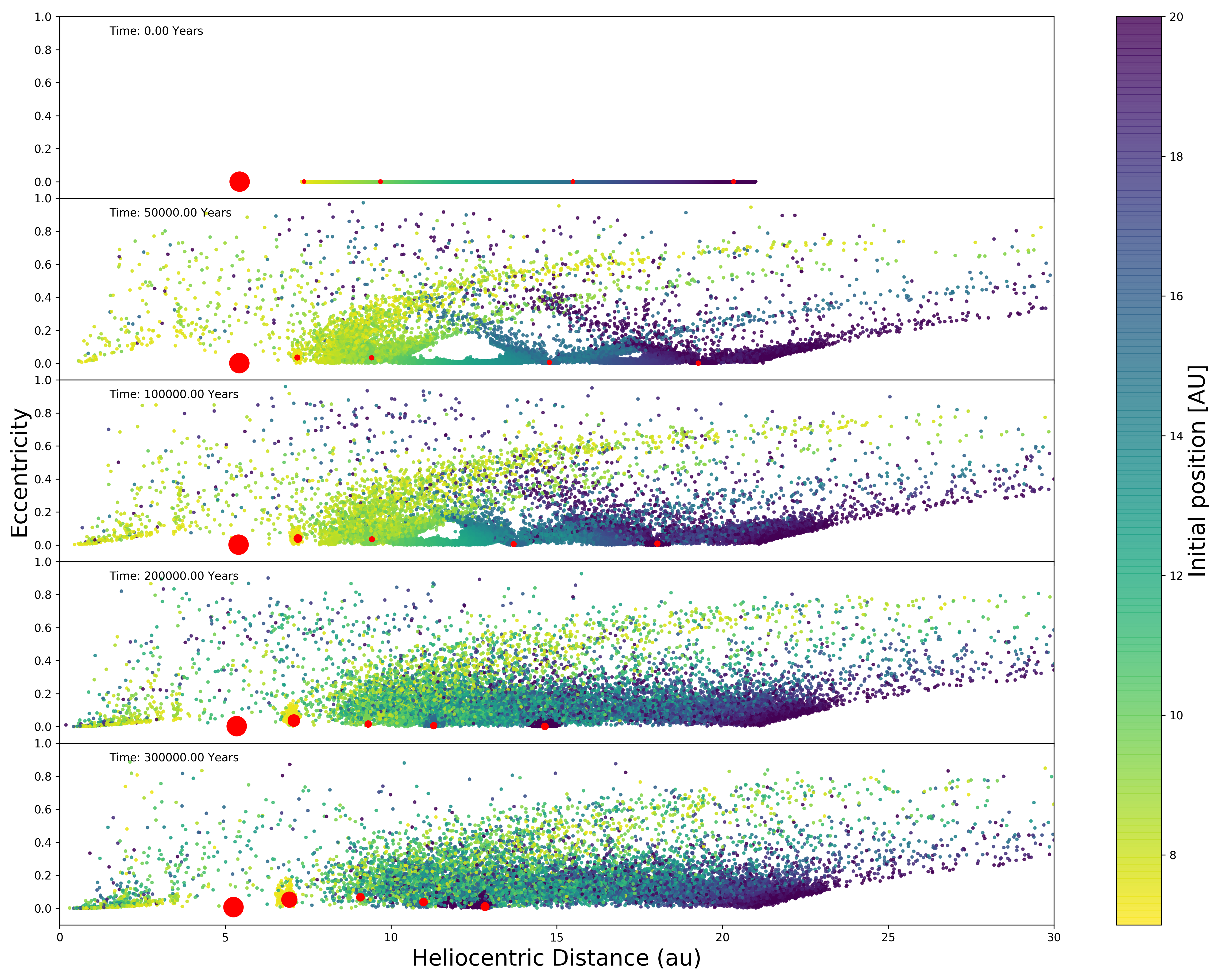}
\caption{
{\bf Migration of ice giants contributes significantly to the implantation of planetesimals into the asteroid belt.}
The temporal evolution of our five-planet system (red) and 20000 planetesimals (multicolored). The heliocentric distance versus eccentricity is shown at the simulation time 0, 0.5, 1, 2, and $3 \times 10^5$ years. The colours of planetesimals correspond to their original locations.
The initial configuration of protoplanets was `wide' (see Tab.~\ref{tab:results}). The final configuration of the solar system will be attained later, during or after gas dispersal. The canonical gas profile (with the surface density $\Sigma_{\rm g} = 300\,{\rm g}/{\rm cm}^2$ at 1\,au) was assumed here. While the growth of the ice giants certainly contributes to the implantation into the asteroid belt, only migration allows for the 1.1:1 CI/CM ratio observed today.
}
\label{fig:timevo}
\end{figure}

\begin{figure}
    \centering
    \begin{tabular}{@{}c@{}}
    \kern.5cm Observed asteroids ($D > 100\,{\rm km}$) \\
    \includegraphics[width=0.5\textwidth]{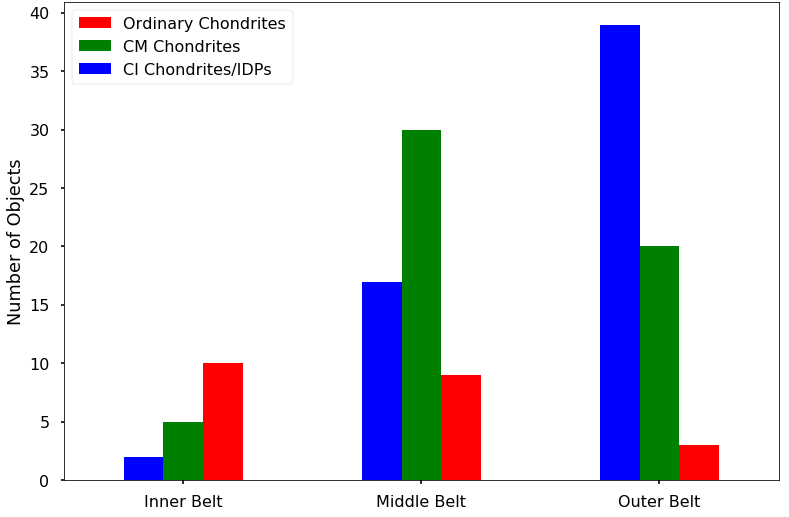} \\
    Migration\kern 4.5cm No migration \\
    \includegraphics[width=\textwidth]{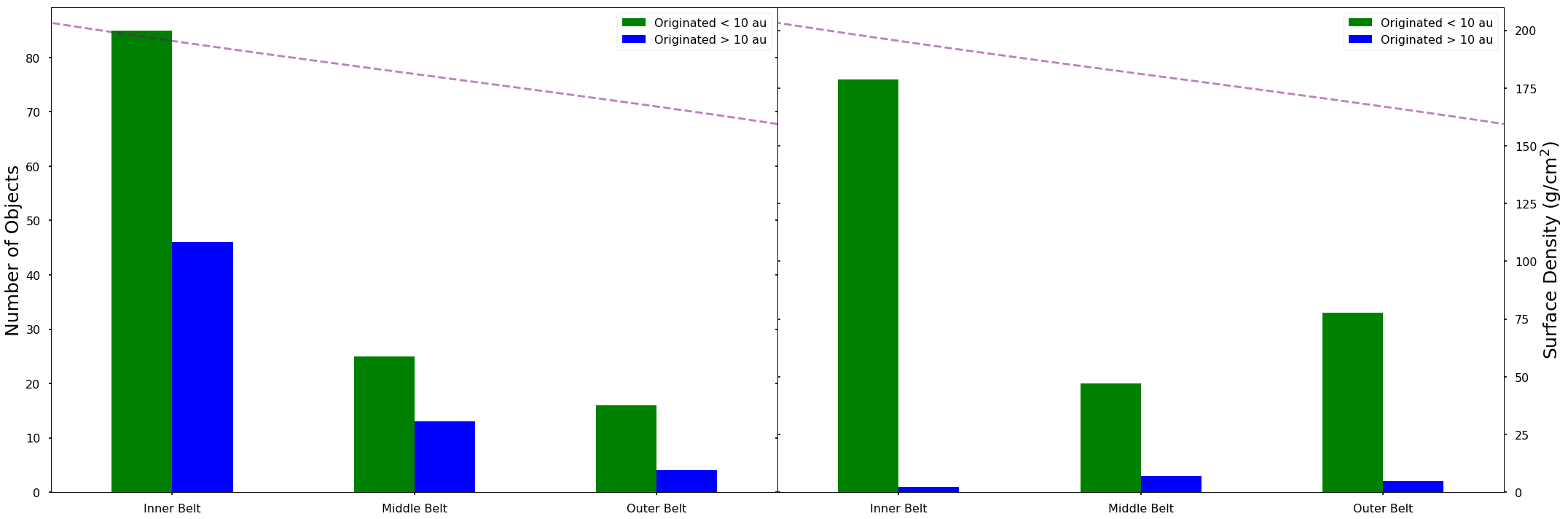} \\
    Gaussian at 2.8\,au\kern 3cm Gaussian at 3.0\,au \\
    \includegraphics[width=\textwidth]{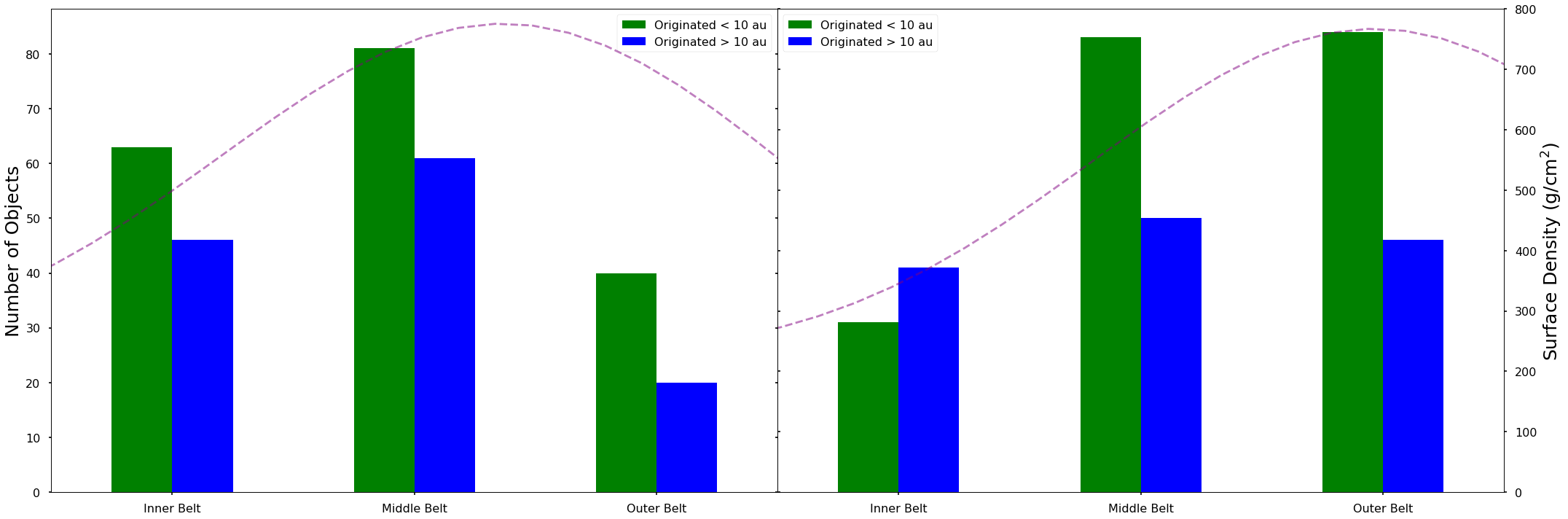} \\
    Desch (3\,Myr)\kern 3.5cm Desch (5\,Myr) \\
    \includegraphics[width=\textwidth]{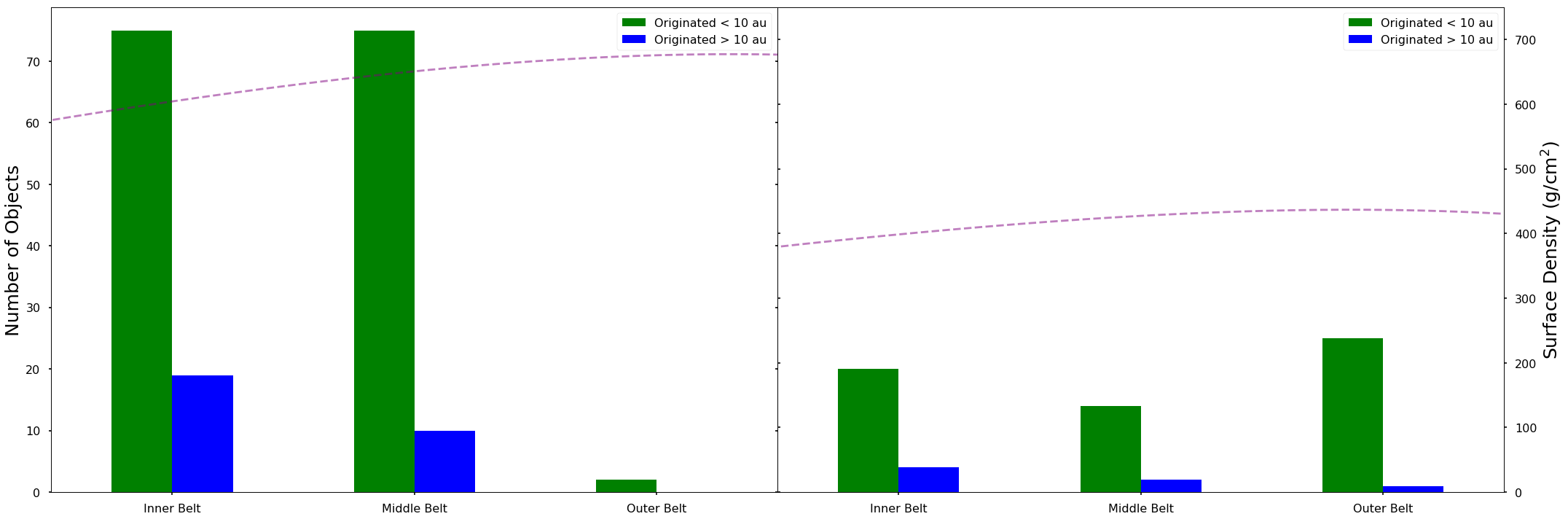} \\
    \end{tabular}
    \caption{
    {\bf Implantation of CM and CI/IDP planetesimals occurred at different times, following the evolution of the gas profile.}
    The number of planetesimals implanted in the inner, middle, and outer belt (solid lines), along with the corresponding gas surface density $\Sigma_{\rm g}(r)$ (dashed lines). In green, the objects successfully implanted in the belt that originally formed at ${<}10$\,au, and in blue, those which formed at ${>}10$\,au. 
    Above: The observed distributions of ordinary, CM-, and CI-chondrite-like asteroids with $D>100$\,km in the belt.
    Top: The canonical model; assuming a wide configuration of planets. The growth timescale of the ice giant planets was $5 \times 10^5\mb{\,{\rm yr}}$. At left, the planetary embryos are allowed to migrate inwards due to their interactions with gas.
    At right, we neglected the interaction with gas and allowed the embryos to grow without migration. 
    Middle: The model with a Gaussian peak; assuming a tight configuration of planets. At left, we changed the gas profile to ensure the peak is located at 2.8\,au. At right, at 3.0\,au.
    Bottom: Model with a wide configuration of planets, using the 3\,Myr gas profile (left) and 5\,Myr gas profile (right) from \citep{Desch2018}. The implantation from the Saturn region is always more efficient than from the trans-Uranian region, but both populations end up in near identical distributions, as dictated by the gas profile. This implies that CI/IDPs were implanted at a later time than CMs; possibly during a gas-free stage.
    }
    \label{fig:mig}
\end{figure}

In the absence of giant planet migration, the vast majority ($>90\%$, see Tab.~\ref{tab:results}) of small bodies captured in the asteroid belt originate from the Saturn region (defined as $a_\text{ini}<10$ au), regardless of the positioning or growth timescales of the ice giants. In all simulations where the gas surface density exceeds $\sim$10\,g/cm$^2$ in the 10-20\,au heliocentric range, however, the interaction between ice giant embryos and the disk gas inevitably leads to their inward migration, dictated by angular momentum exchange with the protoplanetary disk \citep{Ward1997, Izidoro2015}. As a result, we have a migration component in our simulations, most prominent in the `wide' planetary configuration (Fig.~\ref{fig:timevo}): Neptune and Uranus spiral inwards as they grow, in some cases reaching almost 12\,au. This is most notable in rapid planet formation scenarios ($\tau_\text{growth} = 1 \times 10^5$\,yr). The migrating nascent Uranus and Neptune significantly influence the redistribution of planetesimals towards the asteroid belt, to the extent that, at 300\,kyr, a little over half of our implanted objects originate from the ${>}10$\,au region. This underscores that the most efficient driver of the implantation process from the ice giant region is not planetary growth, but migration. Despite their varied origins, the final distribution of objects in the asteroid belt is almost identical to that of bodies originating from the Saturn region (Fig.~\ref{fig:mig}), with a slight trend to lower heliocentric distance. \\

Using a Kolmogorov-Smirnov (KS) test, we compared the distributions of each of our final populations with the observed distributions of CM-like bodies by normalizing the cumulative distributions of the semi-major axis at the 3.2 au gap in the inner belt (Fig.~\ref{fig:mig}). Among our simulations based on synthetic gas profiles, the best-fitting simulation is the one where we forced the gas profile to include a Gaussian pressure bump at a heliocentric distance of 2.7, 2.8, or alternatively 3.0\,au. This would seem to indicate that there was a pressure bump of some kind acting as a planetesimal trap in the vicinity of ${\sim}2.8$\,au during the CM implantation period, i.e., Saturn's growth period. We can think of two mechanisms that could create a pressure bump: The first is a result of the buildup around the water ice line, which was in this vicinity at ${\approx}1$\,Myr after CAIs \citep{Schneeberger2023} or a little further out \citep{Izidoro2022}. However, thermal evolution models imply that CM chondrites formed 3-4\,Myrs after CAIs \citep{Neveu2019}, at a time when the water ice line had been in the terrestrial region \citep{Aguichine2022}. Alternatively, this pressure bump could also simply be the interior edge of Jupiter's gap: if this was the case, the distribution of CMs represents a snapshot of Jupiter's gap at the moment they were implanted in the belt. Finally, none of the gas profiles retrieved from the literature are able to reproduce the asymmetric distribution of CI-like bodies. We ran KS tests also on the cumulative distributions of eccentricities and inclinations, both favored gas profiles with Gaussian pressure bumps, however, these elements were usually damped (as in refs. \cite{Raymond2017a, Nesvorny2024}), which strongly suggests an early gas dispersal or an additional excitation mechanism (see, e.g., \cite{Nesvorn2018}).\\



Overall, our numerical simulations imply that the distribution of small bodies successfully implanted in the asteroid belt mirrors the gas distribution profile of the protoplanetary disk at that time (see Fig.~\ref{fig:gas}). The direct consequence is that the parent bodies of CM and CI chondrites were not implanted into the asteroid belt simultaneously: If they had been, we would expect to observe similar distributions, yet they are markedly distinct (Fig.~\ref{fig:CMCI}). Contrary to previous beliefs, the original formation location of small bodies (e.g., 10\,au versus 20\,au) plays a minimal role \textemdash~per se \textemdash~ in the final distribution in the asteroid belt, for a given ejection time. Rather, the time when they left their birthplace for the asteroid belt is the determining factor. \\

In the absence of precise knowledge of the actual gas profile of the protosolar disk in the 3-10 Myr time window after CAIs, and its evolution over such interval, we need to rely on other modeling and observational constraints to determine the most likely formation locations of CM and CI chondrites. These include but are not limited to:
i)~the current distribution of CM, CI and comet-like (P-type) bodies in the inner solar system (Fig.~\ref{fig:CMCI});
ii)~the current 1.1:1 CI/CM ratio in the asteroid belt (Fig.~\ref{fig:CMCI});
iii)~the similar building blocks for CI chondrites and cometary material such as IDPs ($\sim$100\% matrix, $\sim$0\% chondrules, $\sim$0\% CAIs; \citep{Hutchison2004}); iv)~the comet-like spectral properties of anhydrous areas in CI chondrites \citep{Brunetto2023};
v)~the formation time of CM chondrites (3-4\,Myrs after CAIs \citep{Neveu2019}) and comet-like bodies and TNOs ($\geq$ 4-5\,Myrs after CAIs, \citep{Davidsson2016,Neveu2019, Bierson2019});
vi)~the similarity in spectral properties between P/D-type asteroids, Jupiter Trojans, comets and Centaurs \citep{Vernazza2017};
vii)~the likely late formation of Uranus and Neptune (4-6\,Myrs after CAIs; \citep{Dodson2010, Helled2014});
viii)~the success of the Nice model in explaining many of the observed properties of the solar system (the secular tilt resonance within the cold Edgeworth-Kuiper belt observed today \citep{Baguet2019}, our current planetary configuration, and the capture of Jupiter Trojans and irregular satellites \citep{Nesvorn2018}), including a trans-Neptunian origin for P/D-type asteroids and Jupiter Trojans;
ix)~the asymmetric distribution of bodies implanted in the asteroid belt following the Nice model, a common property with the present distribution of CI and P-type bodies \citep{Vokrouhlicky2016}; and
x)~the size distribution of TNOs which goes up to D$\simeq$2400\,km (Pluto) instead of D$\simeq$270\,km (diameter of the largest inner solar system P/D-type; \citep{Vernazza2021}) making it highly unlikely that no D$\geq$300\,km bodies were trapped in the asteroid belt following the outward migration of Uranus and Neptune as described in the Nice model.
Yet, the Nice model considers only P and D-type asteroids as former TNOs \citep{Levison2009,Vokrouhlicky2016}.
Considering CI-like bodies along with P/D-type asteroids as implanted TNOs would help to solve this serious issue \citep{Vernazza2021}. \\

\begin{figure}
    \centering
    \includegraphics[width=\textwidth]{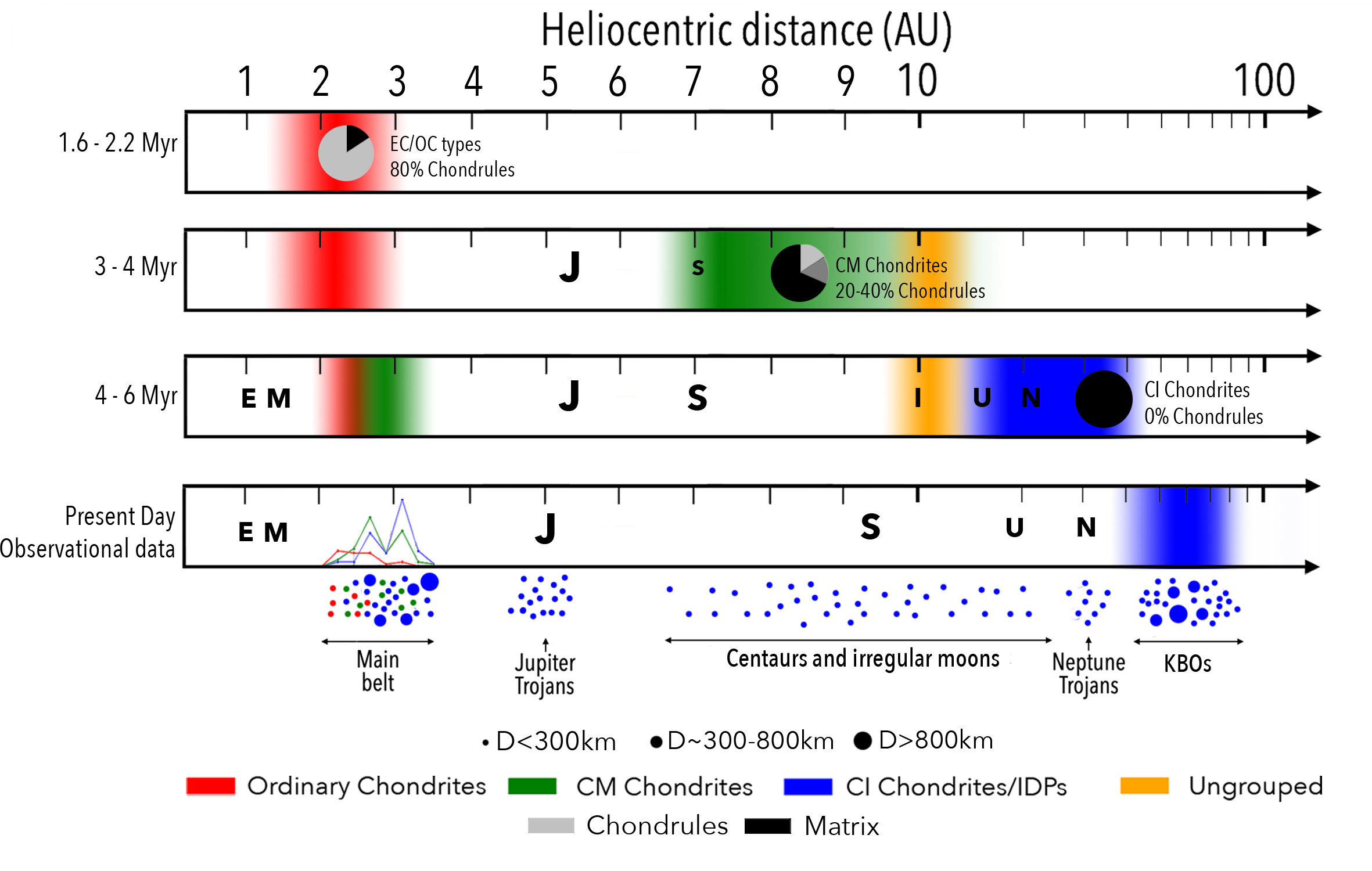}
    \caption{
    {\bf Illustration of the heliocentric distribution of different chondritic materials and their dynamical evolution over time.}
    Three important stages can be distinguished:
    1.6-2.2 Myr (after the formation of CAIs),
    3-4 Myr, and
    4-6 Myr,
    showcasing the temporal and spatial spread of
    Ordinary Chondrites (red),
    CM Chondrites (green),
    and CI Chondrites (blue).
    Pie charts reflect chondrule content of these bodies.
    The bottom panel compares the present-day observational data of the asteroid belt
    and of other populations such as 
    Jupiter Trojans,
    Centaurs,
    Neptune Trojans,d and
    Kuiper Belt Objects (KBOs).
    The circle diameter denotes their typical size.
    Our model explains the implantation of CM chondrite planetesimals from the Saturn region.
    }
    \label{fig:schema}
\end{figure}

From this set of constraints, synthesized in Fig.~\ref{fig:schema} for clarity, there is only one satisfactory solution regarding the formation locations of CM and CI chondrites, namely the Saturn formation region for CM chondrites and the region beyond Uranus and up to Neptune's current position (30\,au) for CI chondrites. CMs were likely implanted during the phase of Saturn's growth at the inner edge of Jupiter's gap. The current distribution of CM-like bodies in the asteroid belt would thus represent the gas profile that was present at the time of this implantation process, and give us a likely indication of the location of Jupiter's gap, a pressure bump in the gas profile being required to reproduce the current CM distribution. CI and IDP-like P/D-types were implanted at a later stage, following the final outward migration of Uranus and Neptune \citep{Tsiganis2005,Morbidelli2005, Gomes2005}, a period characterized by a significantly depleted gas profile, potentially even after gas dispersal. The absence of gas in the disk would lead to an asymmetric distribution of implanted bodies, akin to that of CI-like and P-type bodies \citep{Vokrouhlicky2016} (Fig.~\ref{fig:CMCI}). \\

A challenge remains in reproducing the currently similar distributions in eccentricity and inclination (both in excited states) of CI and CM-like bodies (Fig.~\ref{fig:timevo}). Note that this property can be extended to other taxonomic classes, S-type asteroids in particular. In contrast, our simulations resulted in a distribution pattern that is notably more orderly and less dynamically excited. Additionally, our simulations find themselves with too much mass outward ($\sim$3.5\,au) of Jupiter's 2:1 resonance, compared to what is observed. These two trends are a recurrent issue for all simulations performed at a time when gas is still present in the disk \citep{Raymond2017a, Raymond2017, Raymond2022, Nesvorn2024}. This discrepancy between our simulations and current observational data suggests the presence of an additional, post-implantation dynamical process. Such an event could well be the jump of Jupiter from 5.4 to 5.2 au \citep{Morbidelli2010}, during the final outward migration of Uranus and Neptune (cf. Nice model). The jump would have excited all objects in the belt in the same way while drastically reducing the number of objects beyond Jupiter's 2:1 resonance.\\

Our study has important implications regarding the origin of chondrules (once molten, spherical silicate droplets with diameters of 0.2-2\,mm), which are an essential component of most undifferentiated meteorite classes. Currently, there are two main theories for their origin, namely a nebular origin or a planetary origin (e.g., \citep{Hutchison2004}). Constraining their birthplace would be a major step forward towards deciphering the main physical process responsible for their formation. Here, we have shown that the parent bodies of chondrule-poor CM chondrites (20-40\% chondrules in volume; \citep{Hutchison2004}) likely formed in the Saturn formation region, whereas the parent bodies of chondrule-free CI chondrites (0\% chondrules) likely formed beyond Uranus. Considering that the parent bodies of chondrule-rich enstatite chondrites (ECs) and ordinary chondrites (OCs) ($\geq$80\% chondrules in volume; \citep{Hutchison2004}) certainly formed inward of Jupiter, including in the terrestrial region (in the case of ECs), this implies that chondrule formation occurred mostly inward of the ice giant's formation zone ($\leq$10\,au), with increased efficiency (in volume \% w.r.t. matrix) towards smaller heliocentric distances. This constraint will have to be reproduced in future models aiming to explain the formation process of chondrules. \\

Finally, our most successful simulations
regarding the current distribution of CM-like bodies in the asteroid belt
imply that up to 1.5\% of CM-like bodies from the source region
should have ended up in the formation region (0.4-2\,au) of the terrestrial planets.
This is an unavoidable consequence of their arrival in the asteroid belt.
If the expected mass of planetesimals between Saturn and Uranus is of the order of
${\sim}1\,M_\oplus$ (see Methods),
the average water content of CM-like materials is $10\%$ \cite{Bates2024},
the amount of water implanted among terrestrial planets should be
${\sim}10^{-3}\,M_\oplus$.
This is compatible with the Earth ocean, $2.3\times 10^{-4}\,M_\oplus$,
including also a non-negligible water content in the mantle
(2–8 Earth oceans; ref.~\cite{Peslier_2017SSRv..212..743P}).
Moreover, the D/H ratio of the standard mean ocean water
(i.e., $1.5\times 10^{-4}$; \cite{Craig_1961Sci...133.1702C})
is compatible with a CM-like reservoir \cite{Morbidelli_2000M&PS...35.1309M}.
It follows that CM-like bodies were the main exogenous source of water to Earth and the telluric planets.


\vfill\eject

\vfill\eject

\vfill\eject

\vfill\eject

\begin{table}[!ht]
    \caption{
    Initial conditions and results of 24 of our simulations after 300\,kyr. The `wide' planetary configuration corresponds to Jupiter at 5.4\,au, Saturn at 7.3\,au,  an additional ice giant at 9.7 au \citep{Nesvorn2015, Deienno2017}, Uranus at 15.4 au, and Neptune at\,20.3 au. The `tight' configuration places Uranus at 12.2\,au and Neptune at 16.2\,au. We used a $\sigma$ of 0.2 au for our narrow Gaussian profiles, while those marked with $^*$ used a $\sigma$ of 0.4 au. For each simulation, percentages of planetesimals implanted and the corresponding efficiencies were determined.
    }
    \label{tab:results}
    \centering
    \begin{tabular*}{\textwidth}{@{\extracolsep\fill}lcccccc}
       \toprule
       \small \thead{Gas Profile} & \thead{Growth \\ Timescale \\ (yr)} & \thead{Viscosity \\ $\alpha$ } & \thead{Planet \\ Orbits} & \thead{Implanted \\ by Saturn \\ (\%) } & \thead{Saturn \\ Efficiency \\ (\%) } & \thead{Ice-Giant \\ Efficiency \\ (\%)} \\
       \midrule
       300 g\,cm$^{-2}$ at 1 au & 5$\times 10^{5}$ & 2$\times 10^{-3}$ & Wide    & 64 & 2.13 & 0.30 \\
       300 g\,cm$^{-2}$ at 1 au & 1$\times 10^{6}$ & 2$\times 10^{-3}$ & Wide    & 80 & 1.63 & 0.10 \\ 
       300 g\,cm$^{-2}$ at 1 au & 1$\times 10^{5}$ & 2$\times 10^{-3}$ & Wide    & 55 & 2.50 & 0.51 \\
       300 g\,cm$^{-2}$ at 1 au & 5$\times 10^{5}$ & 2$\times 10^{-3}$ & Tight   & 90 & 1.79 & 0.08 \\ 
       300 g\,cm$^{-2}$ at 1 au & 1$\times 10^{6}$ & 2$\times 10^{-3}$ & Tight   & 74 & 1.88 & 0.17 \\ 
       300 g\,cm$^{-2}$ at 1 au & 1$\times 10^{5}$ & 2$\times 10^{-3}$ & Tight   & 55 & 2.27 & 0.45 \\ 
       300 g\,cm$^{-2}$ at 1 au & 5$\times 10^{5}$ & 2$\times 10^{-3}$ & No Mig. & 93 & 2.45 & 0.05 \\
       300 g\,cm$^{-2}$ at 1 au & 5$\times 10^{5}$ & 5$\times 10^{-4}$ & Wide    & 67 & 2.02 & 0.25 \\
       300 g\,cm$^{-2}$ at 1 au & 5$\times 10^{5}$ & 1$\times 10^{-4}$ & Wide    & 67 & 2.48 & 0.13 \\
       30 g\,cm$^{-2}$ at 1 au  & 5$\times 10^{5}$ & 2$\times 10^{-3}$ & Wide    & 88 & 0.23 & 0.01 \\
       200 g\,cm$^{-2}$ at 1 au & 5$\times 10^{5}$ & 2$\times 10^{-3}$ & Wide    & 76 & 1.57 & 0.13\\ 
       400 g\,cm$^{-2}$ at 1 au & 5$\times 10^{5}$ & 2$\times 10^{-3}$ & Wide    & 61 & 3.20 & 0.51 \\ 
       Gaussian at 2.4 au      & 5$\times 10^{5}$ & 2$\times 10^{-3}$ & Tight   & 75 & 4.13 & 0.60 \\
       Gaussian at 2.7 au      & 5$\times 10^{5}$ & 2$\times 10^{-3}$ & Tight   & 72 & 4.33 & 0.71 \\ 
       Gaussian$^*$ at 2.8 au  & 5$\times 10^{5}$ & 2$\times 10^{-3}$ & Tight   & 60 & 5.54 & 1.55 \\
       Gaussian at 3.0 au      & 5$\times 10^{5}$ & 2$\times 10^{-3}$ & Tight   & 76 & 5.10 & 0.67 \\ 
       Gaussian$^*$ at 3.0 au  & 5$\times 10^{5}$ & 2$\times 10^{-3}$ & Tight   & 62 & 6.06 & 1.59 \\
       Gaussian at 3.2 au      & 5$\times 10^{5}$ & 2$\times 10^{-3}$ & Tight   & 71 & 4.83 & 1.33 \\ 
       Desch 2 Myr             & 5$\times 10^{5}$ & 2$\times 10^{-3}$ & Tight   & 79 & 6.67 & 0.77  \\ 
       Desch 3 Myr & 5$\times 10^{5}$ & 2$\times 10^{-3}$ & Wide    & 86 & 3.79 & 0.11  \\
       Desch 3 Myr        & 5$\times 10^{5}$ & 2$\times 10^{-3}$ & Tight   & 74 & 4.47 & 0.69   \\
       Desch 5 Myr        & 5$\times 10^{5}$ & 2$\times 10^{-3}$ & Tight   & 75 & 3.90 & 0.55    \\ 
       Raymond \& Izidoro     & 5$\times 10^{5}$ & 2$\times 10^{-3}$ & Wide    & 86 & 3.00 & 0.13\\
       Raymond \& Izidoro    & 5$\times 10^{5}$ & 2$\times 10^{-3}$ & Tight    & 56 & 1.50 & 0.50   \\ 
       \bottomrule
    \end{tabular*}
\end{table}


\clearpage

\section*{Methods}\label{sec:methods}


We performed orbital simulations using the open source library \texttt{REBOUND}, a well-established N-body integrator \citep{Rein2012}, selecting the high-accuracy non-symplectic integrator with an adaptive timestepping (IAS15). Each simulation included 20,000 test particles as planetesimals, and the orbits were integrated with a timestep of $10^{-2}/(2\pi)\,\text{yr}$. \texttt{REBOUND} automatically handles collisions of test particles with the planets by removing them from the simulation to save computation time.
Test particles are also removed if their trajectory becomes hyperbolic.  


\subsection*{Giant planet dynamics and growth}

As Jupiter reached approximately $20\,M_\oplus$, it had started to create a pressure bump, isolate pebbles from itself \citep{Lambrechts2014}, and open a gap in the gaseous disk \citep{Lin1986, Lee2002, Crida2017}. After fully opening the gap, small-body populations became separated and Jupiter's migration slowed down. This slowing allowed for a steady accumulation of pebbles over time as pebbles became trapped at its gap's outer edge \citep{Gonzalez2015}, and subsequently accreted into planetesimals. This area of accumulation likely had an increased dust-to-gas ratio, creating conducive conditions for the emergence of larger planetesimals, either through direct cohesion or gravitational instability. Over time, a considerable planetesimal reservoir would have developed at the edge of this gap, with the planet's immediate surroundings remaining relatively barren of solid material. This reservoir's potential significance is underlined by Kobayashi et al. (2012) \citep{Kobayashi2012}, who postulated that Saturn's core might have formed at Jupiter’s gap's outer boundary, a notion echoed by Lambrechts et al. (2014) as it is now clear that Saturn once orbited closer to Jupiter than its present position \citep{Deienno2017}. With Saturn's subsequent formation at Jupiter's disk boundary, its increasing gravitational influence began to channel objects towards Jupiter and the inner Solar System, while also likely providing the material foundation for the creation of the Galilean moons by adding solids to Jupiter's circumplanetary disk (CPD) \citep{Ronnet2018}. Two prominent populations result from this scattering: those trapped on stable orbits within the asteroid belt and those scattered past the belt toward the terrestrial planets, potentially contributing water to nascent terrestrial planets \citep{Meech2020}. The prevalence of each outcome is influenced by the intensity of the gas drag \citep{Raymond2017}, contingent on two main factors: the volumetric density of the gaseous disk, which naturally decreases over time, and the size of the planetesimals, with smaller ones experiencing a heightened drag effect or headwind. Consequently, enhanced gas drag tends to favor implantation, while diminished drag intensifies scattering \citep{Raymond2022}. 

Our model consists of five planets: Jupiter, Saturn, and three ice giants, including Uranus, and Neptune. An additional ice giant (``Ice~1") is needed to stabilize the system after the instability \citep{Nesvorn2015, Deienno2017}. We also chose to neglect the influence of the inner planets, which, having relatively small orbits, require more integration steps and longer calculations.


All planets interact with each other as well as the gas in the disk. Their configuration is dictated by its resemblance to the current Solar System and the observed orbital structure of the trans-Neptunian objects. Our first configuration is `wide', with the planets in a 3:2, 3:2, 2:1, 3:2 multi-resonance configuration following Deienno et al. (2017) \citep{Deienno2017}, as this configuration has been shown to introduce the secular tilt resonance within the cold Edgeworth-Kuiper belt observed today \citep{Baguet2019} as well as many other constraints, such as our current planetary configuration, and the capture of Jupiter Trojans and irregular satellites \citep{Nesvorn2018}. In the latter stages of Solar System formation, as gas depletes from the disk, ``Ice~1" would be expelled during the Jumping Jupiter phase \citep{Morbidelli2010}. Thus, we place Jupiter at a heliocentric distance of approximately $5.4$\,au, consistent with the post-circumsolar disk dynamical evolution of giant planets. We place Saturn's core at the edge of Jupiter's gap, at $7.3$\,au. Following Deienno et al. (2017) \citep{Deienno2017}, ``Ice~1"'s core is placed at 9.7\,au, Uranus's at 15.4\,au, and Neptune's at 20.3\,au. Neptune migrates more or less in the disk as a result of its interaction with the gas. We thus chose to run also simulations with a `tight' configuration, with the cores of Uranus at 12.6\,au, and Neptune at 16.2\,au in a 3:2, 3:2, 3:2, 3:2 resonance chain. We based our configurations on models that have been shown to work once gas had been removed from the disk. In principle, it is possible for these planetary cores to migrate during their formation while gas remains, before assuming the positions that will lead to the Jumping Jupiter configuration.

Jupiter is assumed to have acquired essentially its current mass (${\approx}300\,M_\oplus$) before we even begin our simulations, while the four remaining protoplanets or planetary cores are initialized at $1\,M_\oplus$. Their masses incrementally increase linearly over a parametrized timescale, $\tau_\text{growth}$, as described by:
\begin{equation}
M_\text{planet}(t) = M_i + \Delta M \big(1-{\rm e}^{-t/\tau_\text{growth}}\big)\,,
\end{equation}
where $M_i$ is the initial mass of the core and $\Delta M$ is the difference between the initial core mass and the final mass of each planet. This approach from Ronnet et al. (2018) \citep{Ronnet2018} simplifies the traditional core accretion model, where an envelope contracts, followed by an accelerating and then decelerating gas accretion as a planet forms a gap in the disk \citep{Pollack1996}. We test different values of $\tau_\text{growth}$ set to $1 \times 10^5$, $5 \times 10^5$, and $1 \times 10^6$, in order to see what effect this may have on the motion of the surrounding planetesimals. We use $\tau_\text{growth} = 5 \times 10^5$ for our canonical model. While a deeper analysis is needed for accurate growth patterns, the growth trajectories and order of formation of the ice giants remain complex and are not the primary focus of this study. 

In each case, we included the eccentricity damping of the giant planets due to interaction with the gas disk using fictitious forces. The effects of aerodynamic drag and eccentricity and semi-major axis damping are included in our simulations following Ronnet et al. (2018) \citep{Ronnet2018}, based on the work of Cresswell \& Nelson (2008) \citep{Cresswell2008}. We included the effects of eccentricity and semi-major axis damping of the planets due to interactions with the gas disk through the following acceleration terms:

\begin{equation}
\mathbf{a}_\text{mig} = -\frac{\mathbf{v}}{\tau_\text{mig}}\,,
\end{equation}

\begin{equation}
\mathbf{a}_\text{e} = -2\frac{(\mathbf{v}\cdot\mathbf{r})\mathbf{r}}{r^2\tau_\text{e}}\,,
\end{equation}

\begin{equation}
\mathbf{a}_\text{i} = -\frac{v_z}{t_i}\mathbf{k}\,,
\end{equation}

\noindent where $\mathbf{v}$ is the velocity vector of the planet, $\mathbf{r}$ is its position vector, $\mathbf{k}$ is the unit vector in the $z$-direction, and $r$ is the distance to the star. In the case of Saturn, ``Ice~1", Uranus, and Neptune, the eccentricity damping timescale $\tau_\text{e}$ was taken to be 0.01 $\tau_\text{mig}$ \citep{Lee2002}. As we did not consider any radial migration of Jupiter, we always used an eccentricity damping timescale of $\tau_\text{e}= 5 \times 10^3$ years and no semi-major axis damping for this planet.  The inclination damping time $t_i$ is given by:

\begin{equation}
t_i = \frac{t_{\text{wave}}}{0.544}\times\bigg[ 1 - 0.33\bigg(\frac{i}{h}\bigg)^2\! + 0.24\bigg(\frac{i}{h}\bigg)^3\! + 0.14\bigg(\frac{e}{h}\bigg)^2\bigg(\frac{i}{h}\bigg) \bigg]   \,,
\end{equation}
with
\begin{equation}
t_{\text{wave}} = \frac{M_*}{m_\text{p}}\frac{M_*}{\Sigma_\text{p}a_\text{p}^2}h^4\Omega_\text{p}^{-1} \,,
\end{equation}

\noindent where $i$~denotes the inclination,
$e$~the eccentricity,
$h \equiv H/r$ the aspect ratio,
$M_*$ the stellar mass,
$m_\text{p}$ the planet mass,
$a_\text{p}$ its semimajor axis, and
$\Omega_\text{p}$ the Kepler frequency.

We neglected possible contribution from heating torque from hot/accreting protoplanets \citep{Benitez_2015Natur.520...63B}, torque modifications at non-zero eccentricity \citep{Fendyke_2014MNRAS.437...96F}, or
eccentricity and inclination forcing by the hot-trail effect \citep{Eklund_2017MNRAS.469..206E,Chrenko_2017A&A...606A.114C,Cornejo_2023MNRAS.523..936C}.

These are simplified prescriptions that do not take into account the actual structure of the disk (cf. \citep{Paardekooper_2010MNRAS.401.1950P,Paardekooper_2011MNRAS.410..293P}), as the purpose of this study is not to investigate the precise migration of the giant planets within the disk, but to study implantation of small bodies at specific locations.


\subsection*{Small body dynamics}

Most of the small bodies of the outer Solar System originated from the region between Jupiter and $\sim$30~au \citep{Gomes2003, Levison2008, Tsiganis2005, Kaib2008}. Here, we are interested in the behavior of objects formed beyond Jupiter. With this in mind, we limit our simulations to planetesimals formed in the $7 < a < a_{\text{Neptune}}+1$\,au range, where $a$ is the initial semi-major axis of our test particles. Our model assigns an equal quantity of planetesimals across each semimajor axis segment, not to mirror the actual, unknown distribution but to ensure comprehensive coverage of potential source regions. Variations in the distribution of planetesimals would not significantly alter the likelihood of their implantation into the asteroid belt \citep{Raymond2017}, as interactions between planetesimals are minimal compared to those between planetesimals and planets or gas. The exact distribution of planetesimals is an interesting problem, as we know from observations of exosystems that a continuous disk at this stage is unlikely, and instead, we would have rings of planetesimals around the star resulting from the pressure bumps formed around icelines and other sublimation lines \citep{Izidoro2022}. 

We accounted for the aerodynamic drag effects on the planetesimals using the methods described by Ronnet et al. (2017, 2018) \citep{Ronnet2017, Ronnet2018}, by applying the following acceleration term:

\begin{equation}
\mathbf{a}_\text{drag} = -\frac{1}{t_\text{s}}(\mathbf{v}-\mathbf{v}_\text{g})\,,
\end{equation}

\noindent where $\mathbf{v}_\text{g}$ is the velocity of the gas. The stopping time $t_\text{s}$ is computed using the prescription of Perets \& Murray-Clay (2011) \citep{Perets2011} and Guillot et al. (2014) \citep{Guillot2014}: 

\begin{equation}
t_\text{s} = \left(\frac{\rho_\text{g} v_{\text{th}}}{\rho_\text{s} R_\text{s}}\text{min}\left[1, \frac{3}{8} \frac{v_{\text{rel}}}{v_{\text{th}}}C_\text{D}\right]\right)^{\!\!-1}\,,
\end{equation}

\noindent where $R_\text{s}$ is the size of the solids, here set to 100\,km, and $\rho_\text{s} = 1\,\text{g}\,\text{cm}^{-3}$ their density. The gas volumetric density $\rho_\text{g}$ is obtained from the surface density $\Sigma_\text{g}$ by assuming hydrostatic equilibrium in the vertical direction, with a disk aspect ratio of $h \equiv H/r = 0.05$ \citep{Cresswell2008}. The relative velocity between the gas and the planetesimal is $v_{\text{rel}}$, while the gas thermal velocity is given by $v_{\text{th}} = \sqrt{8/\pi}c_\text{s}$, where $c_\text{s}$ is the isothermal sound speed. In this context, the isothermal sound speed is defined as $c_\text{s} = \Omega H$, being $\Omega$ the Keplerian orbital frequency, expressed as $\Omega = \sqrt{GM_\odot/r^3}$. The dimensionless drag coefficient $C_\text{D}$ is computed as a function of the Reynolds number ${\rm Re}$ of the flow around the planetesimal \citep{Perets2011}:

\begin{equation}
C_\text{D} = \frac{24}{{\rm Re}}(1+0.27{\rm Re})^{0.43} + 0.47 \left(1 - {\rm e}^{-0.04{\rm Re}^{0.38}}\right)\,,
\end{equation}
\begin{equation}
{\rm Re} \equiv \frac{4R_\text{s}v_{\text{rel}}}{c_\text{g}l_\text{g}}\,.
\end{equation}

Using the mean free path of the gas $l_\text{g}$ from Supulver \& Lin (2000) \citep{Supulver2000} and the size of 100\,km, we find a value of $C_\text{D} = 0.44$, which is then fixed as a constant to save computational time.


\subsection*{Gas profiles}

The selection of the disk gas profile stands as a notably challenging endeavor in our modeling. Hydrodynamical simulations have been extensively performed over the years (e.g., \citep{Nelson2000, Bitsch2015, D’Angelo2003, Johansen2015}), each offering nuanced insights and occasionally contrasting perspectives on the gas distribution in protoplanetary disks. These variances arise from the intricate interplay of multiple processes: viscous heating, radiative cooling, stellar irradiation, radiative transfer, dust sublimation, dust settling, and also planet-disk interactions, to name a few. The resultant plethora of potential profiles and the absence of a consensus compel us to make informed yet cautious assumptions when determining the most appropriate gas density profile for our study.

A common result of simplified hydrodynamical disc simulations is to adopt a fixed aspect ratio $h = 0.05$ and a 1-D surface density profile $\Sigma_\text{g} \propto r^{-0.5}$ \citep{Cresswell2008}, normalized so that $\Sigma_\text{g} = 300$\,g\,cm$^{-2}$ at 1\,au \citep{Ronnet2018}, which corresponds to a moderately evolved disk \citep{Bitsch2015}. We also need to account for the gap opening at Jupiter, which we do by removing a Gaussian centered around Jupiter's orbit with of $\sigma$=1\,au so that the surface density of the gas along Jupiter's orbit is $10^{-1}$ lower than it would be otherwise. The turbulent viscosity of the disk is parametrized via $\nu = \alpha c_\text{s}^2 \Omega_{\rm K}$ \citep{Shakura1973}\, with $\alpha$, the viscous parameter of the disk, typically around $\alpha = 2\times 10^{-3}$ for 'dead zones' of the disk.

We also investigate the gas profiles of Desch et al. \citep{Desch2018} and Raymond \& Izidoro \citep{Raymond2017} as a point of comparison, as their models were also successful in implanting objects into the asteroid belt. 

Desch et al. (2018) \citep{Desch2018} initialized the surface density of the gas using a self-similar profile from Hartmann et al. (1998) \citep{Hartmann1998}:

\begin{equation}
\Sigma_\text{g}(r,t=0) = \frac{(2-\gamma)M_\text{disk}}{2\pi R_1} \left(\frac{r}{R_1}\right)^\gamma {\rm e}^{-\left({\textstyle\frac{r}{R_1}}\right)^{2-\gamma}}\,,
\end{equation}

\noindent with $M_\text{disk} = 0.089\,M_\odot$, $\gamma =15/14$, and $R_1 = 1$\,au. The surface density $\Sigma_\text{g}(r,t)$ evolves according to:

\begin{equation}
\frac{\partial\Sigma_\text{g}}{\partial t} = \frac{1}{2\pi r}\frac{\partial \dot{M}}{\partial r}\,,
\end{equation}

\noindent where the mass flux $\dot{M}$ dependence is determined by:

\begin{equation}
\dot{M} = 6\pi r^\frac{1}{2}\frac{\partial}{\partial r}\left(r^\frac{1}{2}\Sigma_\text{g}\nu\right)\,.
\end{equation}

The gap at Jupiter might have been opened at approximately 3\,au \citep{Kruijer2017}. In our simulations, however, Jupiter's gap is located 5.4\,au, since we are examining the effect of Saturn's growth on the implantation efficiency of planetesimals, and, at the same time, most of the CM population is centered around 2.8\,au, where an inner edge of Jupiter's gap was possibly located.


Raymond \& Izidoro (2017) \citep{Raymond2017} assumed the surface density profile of Morbidelli et al. (2017) \citep{Morbidelli2007}, starting after Jupiter and Saturn were both formed. The starting disk profile has a $\Sigma_\text{g}$ of 200\,g\,cm$^{-2}$ at 1\,au, a gap opened by Jupiter at 5.4\,au, and a partial gap opened around Saturn at 7\,au. While they decreased the surface density uniformly in radius, on a $2 \times 10^5$\,yr exponential timescale, and removed entirely after $2 \times 10^6$\, yr, we chose to examine their starting disk profile, as the gap opened by Jupiter and Saturn was already much wider than in our profiles. 

\begin{figure}
\centering
\includegraphics[width=\textwidth]{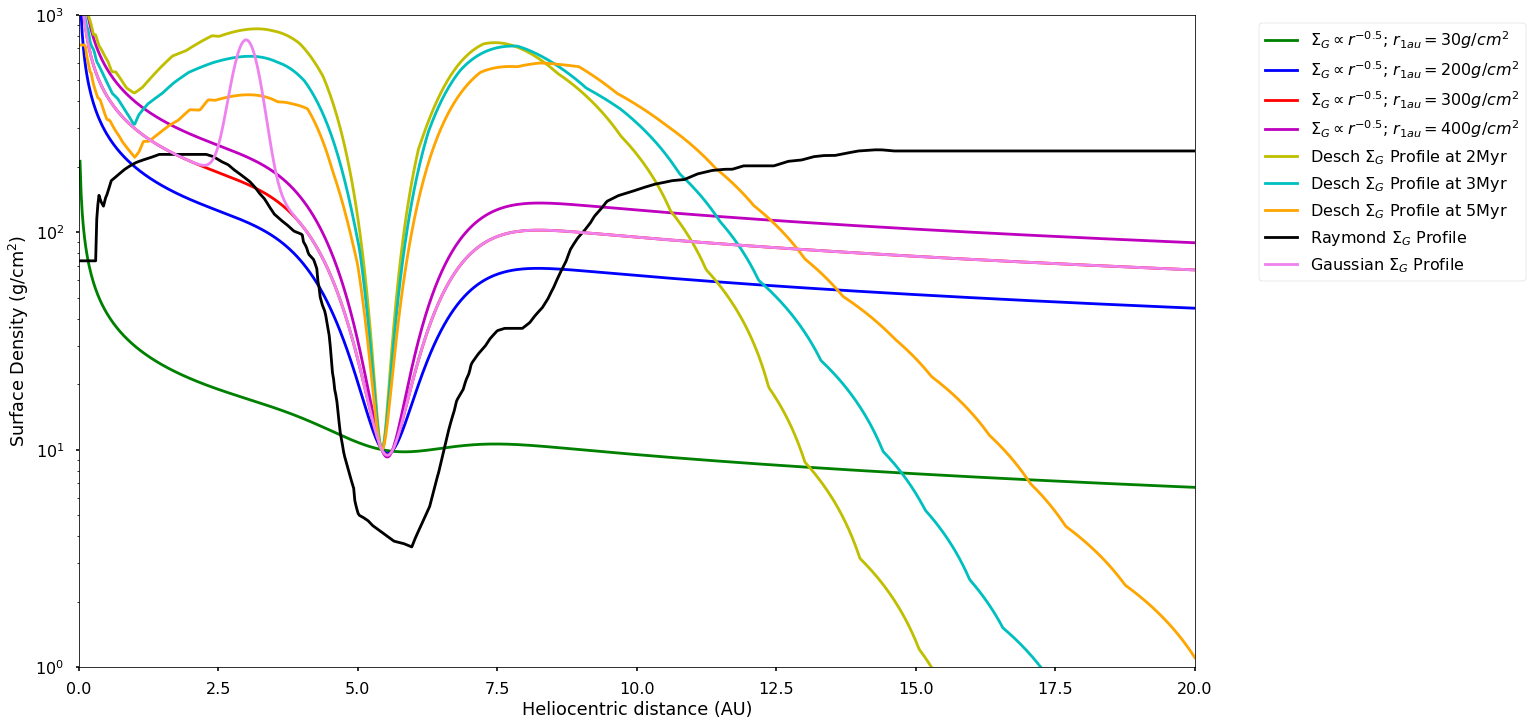}
\caption{
{\bf A comparison of the different disk profiles used throughout this study.}
The surface density $\Sigma_\text{g}(r)$ is plotted as a function of the heliocentric distance~$r$.
We assumed Jupiter has opened a gap in the gaseous disk at approximately 5.4\,au.
Its current position 5.2\,au has been attained at later times, during a gas-free stage.
The exact shape of this gap is not very important.
We examined the canonical profile, a Gaussian peak approximation (which is valid around 3\,au), the Desch \citep{Desch2018} profile at different periods, and the initial Raymond \citep{Raymond2017} profile.}
\label{SigmaG}
\end{figure}

A comparison of all these gas surface density profiles is shown in Fig.~\ref{SigmaG}.


\subsection*{Mass of planetesimals disk}

The total mass of solids available to form planetesimals and protoplanets
is varied among our models.
In particular, we were concerned with the mass between Saturn and Uranus,
which corresponds to CM-like material.
The upper limit is given by integrating the canonical gas profile
($\Sigma_\text{g}$ of 300\,g\,cm$^{-2}$ at 1\,au)
over 7.3 to 15.4\,au,
in the case of a `wide' configuration of planets.
For an assumed metallicity
(0.02; based on ref. \citep{VonSteiger2016})
this amounts up to $38\,M_\oplus$.
However, a substantial part of this material contributes to the growth of
``Ice 1", Uranus, possibly also Saturn's core.
Since we do not model the growth itself (only parametrically),
we should subtract at least
14.5 and $14.5\,M_\oplus$,
so that the remaining planetesimals amount to approximately $9\,M_\oplus$.
The lower limit is given by assuming
a `tight' configuration of planets (12.6\,au),
a lower metallicity (0.01).
This would lead to $12\,M_\oplus$,
which seems to be insufficient for the respective planets,
nevertheless, we can conclude that the lower limit
for the mass of planetesimal disk
is of the order of ${\approx}1\,M_\oplus$.

\subsection*{Arguments against implantation of CI-like bodies}

Our simulations revealed an excessive efficiency of implantation for CI bodies. Supposing the total mass of planetesimals in the trans-Uranus region is around $30\,M_\oplus$ \textemdash~ as is required to drive Neptune's outward migration \cite{Nesvorny_2018ARA&A..56..137N} \textemdash~ 5\% would be aerodynamically implanted in the belt.
Even if late instabilities depleted this by a factor of 100, the remaining $0.015\,M_\oplus$ still exceeds the current asteroid belt mass of $4\times 10^{-4}\,M_\oplus$ \citep{Pitjeva_2018AstL...44..554P}.

Similarly, about 1.5\% of CI-like bodies, or $0.45\,M_\oplus$,
would be implanted in the terrestrial region.
We counted all bodies located among terrestrial planets at 300\,ky, given that all orbits have undergone significant damping at this stage. These will be accreted by the terrestrial planets in due course.
With an estimated 10\% water content \cite{Bates2024}, the resulting $0.045\,M_\oplus$ still exceeds the upper limits for the Earth mantle’s water content, up to $2\times 10^{-3}\,M_\oplus$ \cite{Peslier_2017SSRv..212..743P}.

The only solution to this problem is that CI-like bodies were {\em not\/} implanted during the gas phase, as in ref. \cite{Nesvorny2024},
but after gas dispersal, with a much lower implantation efficiency
of the order of $4\times 10^{-6}$ \cite{Vokrouhlicky_2016AJ....152...39V}.
This results in the correct mass of CI-like bodies (i.e., a quarter of the asteroid belt), and a negligible contribution to the terrestrial planets.


\backmatter






\vskip2\baselineskip

\bmhead{Acknowledgments}

The work of M.B. has been supported by the Czech Science Foundation (grant number 21-11058S).

\bmhead{Author contributions}

The authors contributed equally to this work.

\bmhead{Competing interests}

The authors declare no competing interests.


\vfill\eject

\vfill\eject

\begin{figure}
    \centering
    \begin{tabular}{@{}c@{}}
    Canonical profile ($\Sigma_\text{g} = 300\,{\rm g}\,{\rm cm}^{-2}$ at 1\,au) \\
    \includegraphics[width=\textwidth]{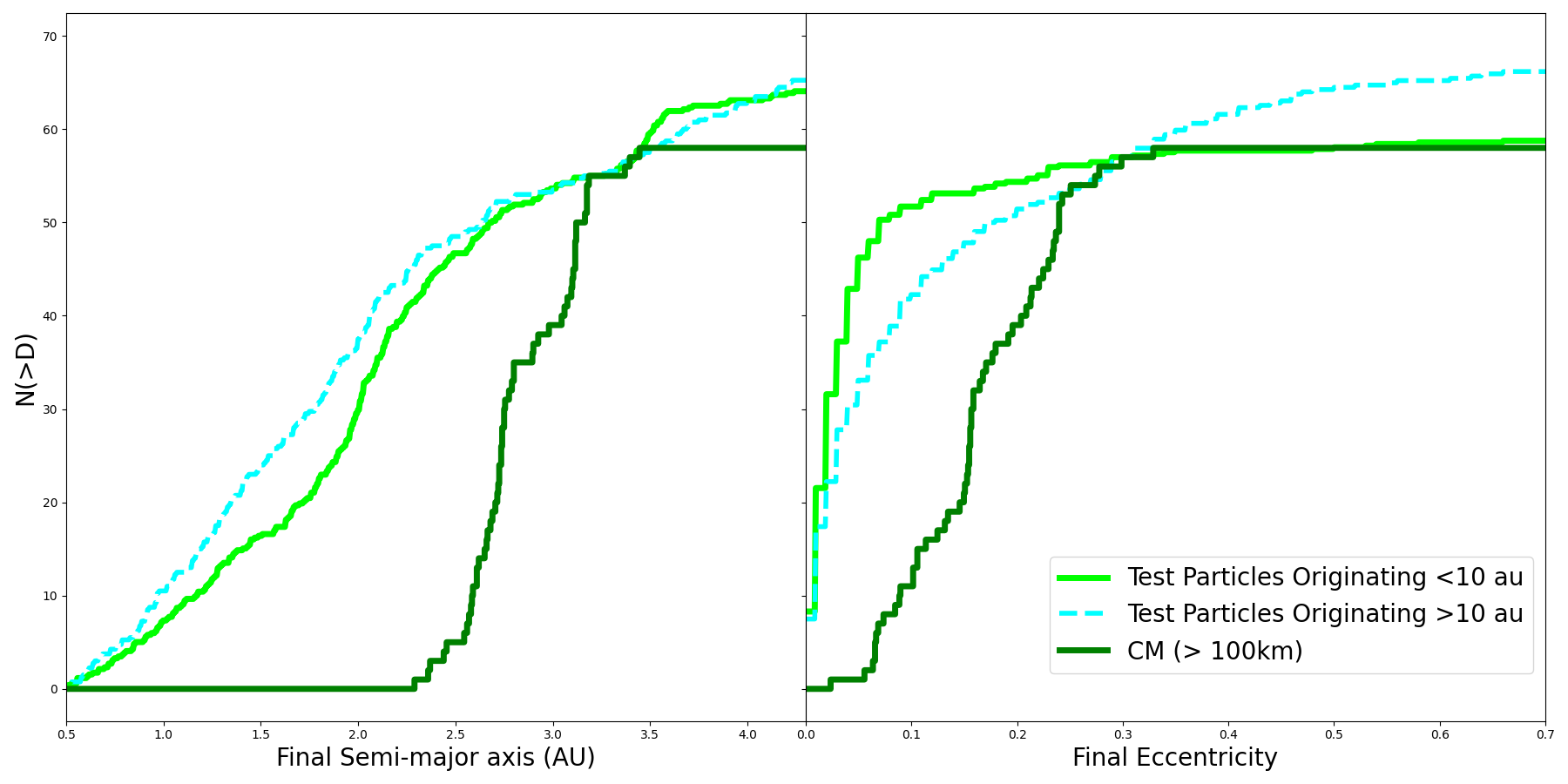} \\
    Gaussian at 3\,au \\
    \includegraphics[width=\textwidth]{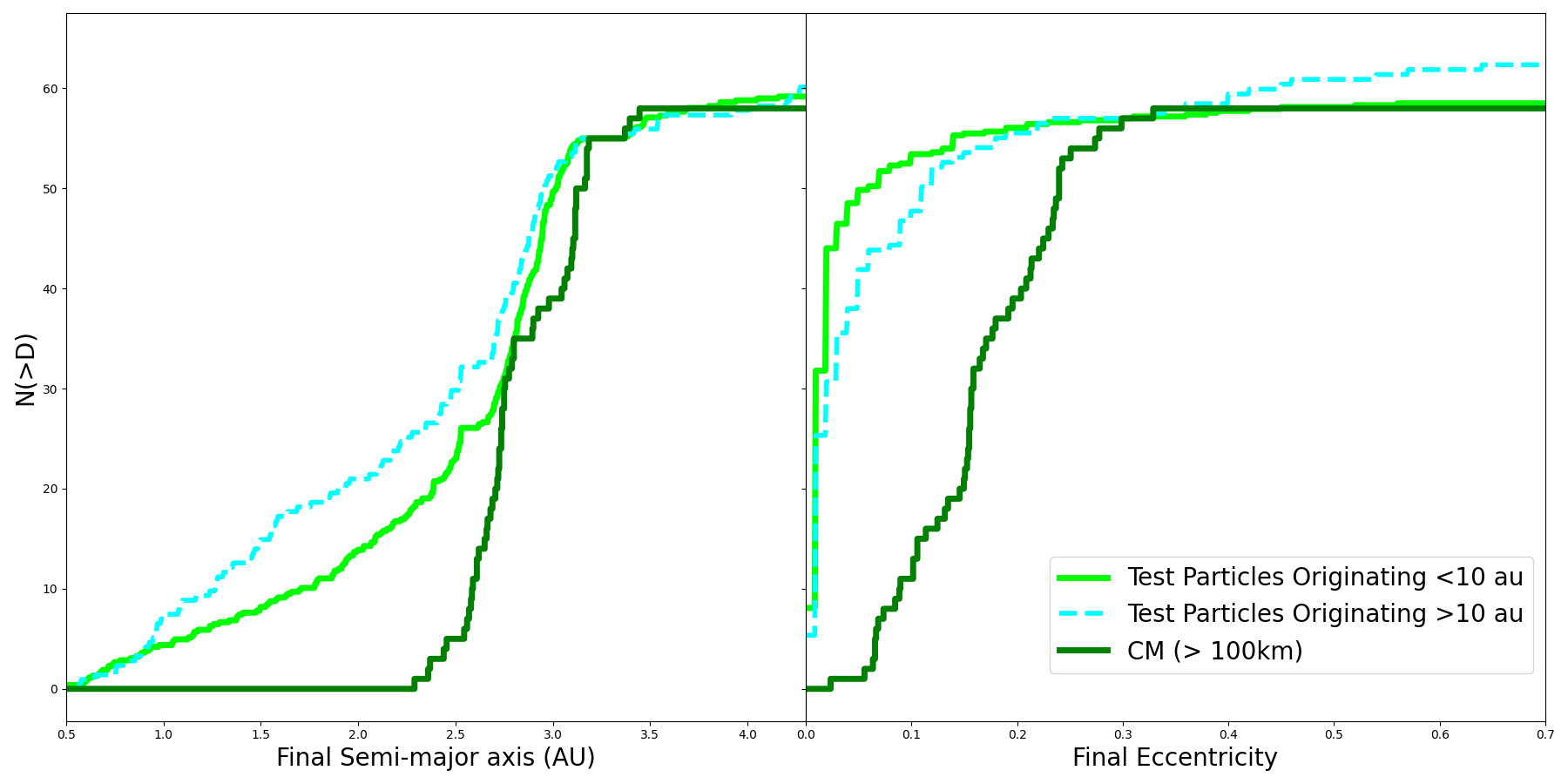} \\
    \end{tabular}
    \caption{
    \textbf{A pressure trap in the gas at 2.7 to 3\,au would allow planetesimals
    arriving from Saturn to reach their observed radial distribution.} 
    Cumulative distributions of semi-major axes (left) and eccentricities (right)
    of our models at 300\,kyr simulation time
    are compared to the observed distribution of CM- (\textcolor{green}{green}) and CI- (\textcolor{blue}{blue}) chondrites.
    Top: An excluded model with the canonical surface density profile and the tight configuration of planets.
    Bottom: A preferred model with a Gaussian pressure bump placed at 3.0\,au, with a $\sigma$ of 0.4\,au.
    The arriving planetesimals were split by their origin,
    with the black line representing those that formed near Saturn ($<$10\,au).
    The distributions are normalized to the CM distribution at $a=3.2$\,au, $e=0.3$.
    The excess of planetesimals ${<}2$\,au compared to what is observed today
    reveals their contribution to the water content of the terrestrial planets.
    The main difference between our simulations and the observed distributions
    is that the eccentricities (and inclinations) are strongly dampened.
    This results from the presence of gas in the disk; it strongly suggests
    an excitation event occurred after CMs arrived in the belt.
    }
    \label{fig:distrib}
\end{figure}












\begin{appendices}







\end{appendices}



\clearpage

\bibliographystyle{unsrt}
\bibliography{sn-bibliography.bib}


\end{document}